\documentclass[12pt]{article}
\usepackage{fullpage,graphicx,psfrag,amsmath,amsfonts,verbatim}
\usepackage[small,bf]{caption}
\usepackage{booktabs}
\usepackage{graphicx}
\usepackage{subcaption}

\usepackage{hyperref}

\newcommand{\ones}{\mathbf 1}
\newcommand{\reals}{{\mbox{\bf R}}}

\newcommand{\eg}{{\it e.g.}}
\newcommand{\ie}{{\it i.e.}}

\newcommand{\BEAS}{\begin{eqnarray*}}
\newcommand{\EEAS}{\end{eqnarray*}}
\newcommand{\BEA}{\begin{eqnarray}}
\newcommand{\EEA}{\end{eqnarray}}
\newcommand{\BEQ}{\begin{equation}}
\newcommand{\EEQ}{\end{equation}}
\newcommand{\BIT}{\begin{itemize}}
\newcommand{\EIT}{\end{itemize}}

\title{Finding Moving-Band Statistical Arbitrages\\ via
Convex-Concave Optimization}
\author{Kasper Johansson \and Thomas Schmelzer \and Stephen Boyd}

\begin{document}
\maketitle

\begin{abstract}
We propose a new method for finding statistical arbitrages that can contain more
assets than just the traditional pair.  We formulate the problem as seeking 
a portfolio with the highest volatility, subject to its price remaining in a band
and a leverage limit.
This optimization problem is not convex, but can be approximately solved 
using the convex-concave procedure, a specific sequential convex programming method.
We show how the method generalizes to finding
moving-band statistical arbitrages, where the price band midpoint varies over time.
\end{abstract}

\newpage
\tableofcontents
\newpage

\section{Introduction}\label{s-intro} 
We consider the problem of finding a
\emph{statistical arbitrage} (stat-arb), \ie, a portfolio with mean-reverting 
price.
Roughly speaking, this means that the price of the portfolio stays in a band,
and varies over it.
Such a stat-arb is traded in the obvious way, buying when the price is in
the low part of the band and selling when it is in the high part of the band.

Traditional stat-arbs focus on portfolios consisting of two or possibly three
underlying assets.  When the portfolio contains two assets, trading the stat-arb
is called \emph{pairs trading}.  Pairs can be found by exhaustive search over
all $n(n-1)/2$ pairs of assets in a universe of $n$ assets.
When the weights of the two assets in pairs trading are $+1$ and $-1$, the portfolio
value is the \emph{spread} (between the two prices).
The assets in a pair are called \emph{co-moving assets}.

In this paper we propose a new
method for finding stat-arbs that can contain multiple (more than two) assets,
with general weights.  The problem is
formulated as a nonconvex optimization problem in which we maximize the
portfolio price variation subject to the price staying within a fixed band,
along with a leverage limit, over some training period.
Although this approach requires maximizing a convex function, 
we show how to approximately
solve it using the convex-concave procedure~\cite{shen2016disciplined,
lipp2016variations}. 

Our second contribution is to introduce the concept of a \emph{moving-band
stat-arb}, in which the price of the portfolio varies in a band that changes over
time, centered at the recent average portfolio price.
(We refer to a traditional stat-arb as a \emph{fixed-band stat-arb}.)
We show that the same method we use to find fixed-band stat-arbs can be 
used to find moving-band stat-arbs, despite the apparent complexity of
the average price also depending on the portfolio.
Moving-band stat-arbs are traded in the same obvious way as fix-band stat-arbs, 
buying when the price is in
the low part of the band and selling when it is in the high part of the band;
but with moving-band stat-arbs, the center of the band changes over time.
Moving-band stat-arb trading resembles trading using \emph{Bollinger
bands}~\cite{bollinger2002bollinger, bollinger1992using}, except that the 
bands are associated with the price of a carefully constructed portfolio,
and not a single asset.
Our empirical studies show that moving-band stat-arbs out-perform fixed-band
stat-arbs in terms of profit,
and remain profitable for longer out-of-sample periods. 

\subsection{Related work}\label{s-related-work}

\paragraph{Stat-arbs.}
Stat-arb trading strategies date back to the 1980s when a group of Morgan
Stanley traders, led by Nunzio Tartaglia, developed pairs trading
\cite{pole2011statistical, gatev2006pairs}. This strategy involves identifying pairs of assets whose price tends
to move together, hence referred to as pairs trading. In pairs trading, the spread,
\ie, price difference between two assets, is tracked and positions entered when this
difference deviates from its mean. This trading strategy has enjoyed widespread
popularity, with its success substantiated by numerous empirical studies in
various markets like equities~\cite{avellaneda2010statistical},
commodities~\cite{nakajima2019expectations, vaitonis2017statistical}, and
currencies~\cite{fischer2019statistical};
see, \eg,~\cite{gatev2006pairs, avellaneda2010statistical, perlin2009evaluation,
hogan2004testing, krauss2017deep, caldeira2013selection, huck2019large, dunis2010statistical}.

In the general setting, a stat-arb consists of multiple assets in a
portfolio that exhibits a mean-reverting behavior~\cite[\S 10.5]{feng2016signal}. Stat-arb trading is a widely
used strategy in quantitative finance. 
The literature on stat-arbs is extensive and generally splits into several
categories:
finding stat-arbs,
modeling the (mean-reverting) portfolio price, and trading stat-arbs.
We give a brief review of these here and refer the reader 
to~\cite{krauss2017statistical} for a
comprehensive overview of the literature.

\paragraph{Finding stat-arbs.}
Probably the simplest approach to finding pairs of co-moving securities is the distance
approach. The distance pairs trading strategy finds assets whose (normalized)
prices have moved closely historically, in an exhaustive search through pairs of
assets~\cite{gatev2006pairs}. Assets whose prices have a low sum of squared
deviation from each other are considered for trading. The
distance approach is simple and intuitive, although it does not necessarily find
good pairs~\cite{krauss2017statistical}. The objective itself is to minimize the
distance between two asset prices, which does not directly relate to the desired
properties of a stat-arb, which crucially should also have a high variance. 
This paper addresses this issue by directly optimizing for
large fluctuations around the mean.

The co-integration approach is another popular method for finding co-moving
securities. Co-integration
is an important concept in the econometrics
literature~\cite{JOHANSEN2000359, alexander2005indexing}, and dates back to Engle and Granger's works
in the 1980s (for which Granger was awarded the 2003 Nobel Memorial Prize in
Economic Sciences)~\cite{granger1983co, engle_granger_1987}. The idea is that
the absence of stationarity in a multivariate time series may be explained by
common trends, which would make it possible to find linear combinations of assets
that are stationary
and hence mean-reverting. Thus, the co-integration approach is based on
identifying linear combinations of assets that result in a stationary
time series~\cite{krauss2017statistical}. In~\cite{vidyamurthy2004pairs}, the most cited work on co-integration based pairs
trading, potential asset pairs are found based on statistical measures, which
are then tested for co-integration using an adapted version of the Engle-Granger
test. Several co-integration based methods have been proposed to extend the pairs
trading strategy to more than two assets. For example, in~\cite{zhao2018mean,
zhao2018optimal, zhao2019optimal} the authors consider a (non-convex)
optimization problem for finding high variance, mean-reverting portfolios. Their strategy is based on finding a portfolio of
spreads, defined by a co-integration subspace, and implemented using sequential
convex optimization. Their proposed optimization problem, \ie, maximizing variance subject to a mean
reversion criterion is similar to our problem formulation. However, our problem
differs significantly in that we do not rely on any co-integration analysis or
statistical testing. Rather, we directly optimize for a high variance portfolio
that is mean-reverting.

Asset pairs can also be found using machine learning methods.
In~\cite{SaHo2020mlpairs}, the authors use unsupervised learning and propose a
density-based clustering algorithms to cluster assets. Then,
within asset clusters, pairs of assets are chosen for trading depending on
co-integration, as well as mean-reversion tendency and frequency. Modern machine
learning methods are also explored in \cite{krauss2017deep}, where the authors propose the use of deep
neural networks, gradient-boosted trees, and random forests for finding stat-arb
portfolios. Another recent study of deep-learning stat-arb finding
is~\cite{guijarro2021deep}. Earlier work on using machine learning for finding
stat-arbs includes, \eg,~\cite{dixon2015, moritz2014deep, takeuchi2013applying,
	huck2010pairs, huck2009pairs}.

\paragraph{Modeling the stat-arb spread.}
When a co-moving set of assets has been identified, the next step is to 
model the portfolio price (or spread between the assets for a pair). Perhaps the
most popular approach is to model the spread using stochastic control theory. It
is common to consider investments in a mean-reverting spread and
a risk-free asset and to model the spread as an Ornstein-Uhlenbeck
process~\cite{mudchanatongsuk2008optimal, jurek2007dynamic}. Other methods
include those borrowing tools from time series
analysis~\cite{krauss2017statistical}. 
In~\cite{elliott2005pairs} the authors propose a mean-reverting
Gaussian Markov chain model for modeling the spread between two assets. Copulas
have also been proposed to model the joint distribution of the spread, both for pairs and for 
larger sets of assets~\cite{de2016pairs, stubinger2018statistical, krauss2017non}.
In~\cite{de2016pairs} the authors suggest modeling the spread using linear state
space models.


\paragraph{Trading stat-arbs.}
We mention here a number of stat-arb trading methods, ranging from simple ones based
on the intuitive idea of buying when the price is low and selling when it is 
high, to more complex ones based on learning the price statistics and using 
stochastic control.
One simple method is based on \emph{hysteresis}, as is used in a
conventional thermostat.  In this approach we buy (enter into a long position) when
the price drops below a threshold, and sell (switch to a short position) when 
the price goes above another threshold.  The thresholds are typically based on 
price bands, as discussed below. A variation on this method sets the thresholds based on the standard
deviation of the price, as proposed in~\cite{gatev2006pairs}. 
Another simple method is \emph{linear trading}, where we take a position
proportional to the difference between the band midpoint price and the current
price.  This method can also be modified to use volatility-based bands
instead of fixed bands.  (Such a trading policy is a simple Markowitz policy,
with the mean return given by the difference between the midpoint price
and the current price.)

Other methods for trading stat-arbs follow directly from the 
spread models described above. When the
spread is modeled as an Ornstein-Uhlenbeck process, the optimal trading strategy
is found by solving a stochastic control
problem~\cite{mudchanatongsuk2008optimal, jurek2007dynamic,
bertram2010analytic}. In \cite{yamada2012model, primbs2018pairs, yamada2018model} the authors model the spread as
an autoregressive process, a discretization of the Ornstein-Uhlenbeck
process, and show how to trade a portfolio of spreads under proportional transaction
costs and gross exposure constraints using model predictive control. With the copula
approaches of~\cite{de2016pairs, stubinger2018statistical, krauss2017non}, the trading strategy is based on deviations from
confidence intervals. The machine learning methods of~\cite{SaHo2020mlpairs,
krauss2017deep, guijarro2021deep, dixon2015, moritz2014deep,
takeuchi2013applying, huck2010pairs, huck2009pairs} also include trading
strategies.

\paragraph{Exiting a stat-arb.} A stat-arb will not keep its mean-reverting
behavior forever. Hence, a strategy for exiting a
stat-arb, \ie, closing the position, is needed. One approach is to exit when the
spread reaches a certain threshold in terms of its standard deviation or in terms
of a price-band, as described below. Another
simple method is to exit after a fixed time-period, as was done
in~\cite{gatev2006pairs}.  

A simple variation on any of these exit methods does not exit the position 
immediately when the exit condition is first satisfied.  
Instead it reduces the position to zero slowly (\eg, linearly) over some 
fixed number of periods.

\paragraph{Price-bands.}
Price-bands are popular in technical analysis, and are used to identify
trading opportunities based on the price of an asset relative to its recent 
price history~\cite{lee2021price}. The most popular is the
Bollinger band~\cite{bollinger1992using}. It is constructed by computing the
$M$-day moving average (with common choice $M=21$) of the asset price, 
denoted $\mu_t$ at time $t$, and the corresponding price standard deviation 
$\sigma_t$. The Bollinger band is then defined as the interval $[\mu_t - k\sigma_t, \mu_t +
k \sigma_t]$, where $k>0$ is a parameter, typically taken to be $k=2$.
A price signal is extracted based on
if the price is near the top or bottom of the band. For a detailed description
of the Bollinger band and other trading-bands, we refer the reader
to~\cite{bollinger2002bollinger,bollinger1992using}.

\subsection{This paper} 
This paper proposes a new method for finding stat-arbs, with two
main contributions. The first contribution of our method is to formulate the
search for stat-arbs as an
optimization problem that intuitively and directly relates to the desired 
properties of a stat-arb: its price should remain in a band (\ie, be mean-reverting)
and also should have a high variance.
Since our method is based on convex optimization, it readily scales to large universes of
assets. It can and does find stat-arbs with ten or more assets, well beyond our ability 
to carry out an exhaustive search.   

Our second contribution is to introduce the concept of a moving-band stat-arb.
While the idea of a price band is widely known and used in technical analysis 
trading, the main difference
is that we apply it to carefully constructed portfolios (\ie, stat-arbs), instead of 
single assets.

Our focus is on finding both fixed-band and moving-band stat-arbs, and not on 
trading them.  Our numerical experiments use a simple linear trading policy, and 
a simple time-based exit condition. (We have also verified that similar results 
are obtained using hysteris-based trading policies.)
We also do not address the question of how one might trade a portfolio of 
stat-arbs, the focus of a forth-coming paper.

\subsection{Outline}
The rest of the paper is organized as follows. In \S\ref{s-finding-statarbs} we
propose a new method for finding traditional stat-arbs, \ie,
with a fixed band. 
We extend this method to moving bands in \S\ref{s-movingband-statarbs}. In
\S\ref{s-experiments} we present experiments on real data, and in
\S\ref{s-conclusion} we conclude the paper.

\section{Finding fixed-band stat-arbs}\label{s-finding-statarbs}
We consider a vector time series of prices of a universe of $n$ assets, denoted $P_t \in
\reals^n$, $t = 1, 2, \ldots, T$, denoted in USD per share. 
(We presume these are adjusted for dividends and splits.)
We consider a portfolio of these assets given
by $s\in \reals^n$, denoted in shares, with $s_i<0$ denoting short positions.
The price or net value of the portfolio is the scalar time series $p_t = s^T
P_t$, $t=1,\ldots, T$. The asset prices $P_t$ are positive, but the portfolio
price $p_t$ need not be, since the entries of $s$ can be negative.
For future use, we define the average price of the $n$
assets as the vector $\bar P= (1/T)\sum_{t=1}^T P_t$.

We seek a portfolio for which $p_t$ consistently varies over a band
(interval of prices) with two goals.
It should stay in the price band,
and also, vary over the band consistently; that is, it should
frequently vary between the high end of the band and the low end of the band.
We refer to such a portfolio $s$ as a stat-arb, so we use the term to 
refer to the general concept as well as a specific portfolio.
Our method differs from the traditional statistical framework, where one 
would seek a portfolio $s$ for which $p_t$ is co-integrated.

\subsection{Formulation as convex-concave problem}\label{s-problem}
We formulate the search for a stat-arb $s$ as an optimization problem. The
condition that $p_t$ varies over a band is formulated as $-1 \leq p_t -\mu \leq 1$, where $\mu$
is the midpoint of the band.  Here we fix the width of the price band as
$2$; $s$ and $\mu$ can always be scaled so this holds. We express the desire that
$p_t$ vary frequently over the band by maximizing its volatility.
Since $s$, $p$, and $\mu$ can all be multiplied by $-1$ without any
effect on the constraints or objective, we can assume that $\mu\geq 0$ without
loss of generality.  

We arrive at the problem 
\BEQ
\begin{array}{ll}
\mbox{maximize} & \sum_{t=2}^T (p_{t} - p_{t-1})^2\\
\mbox{subject to} & -1 \leq p_t - \mu \leq 1, \quad p_t=s^TP_t, \quad t = 1,
\ldots, T\\
& |s|^T \bar P \leq L, \quad \mu \geq 0,
\end{array}
\label{e-stat-arb-prob}
\EEQ 
with variables $s \in \reals^n$, $p \in \reals^T$, and $\mu \in \reals$, where the absolute value in the last constraint is elementwise. The problem data are
the vector price time series $P_t$, $t=1,\ldots, T$, and 
the positive parameter $L$.  

Note that $|s|^T \bar P$ is the average total position of the portfolio, sometimes
called its leverage, so the constraint  $|s|^T \bar P \leq L$ is a leverage
constraint; it limits the total position of the portfolio. 
The leverage is a weighted $\ell_1$ norm of $s$, and so tends to lead
to sparse $s$, \ie, a portfolio that concentrates in a few assets, a typical
desired quality of a stat-arb.
The problem
\eqref{e-stat-arb-prob} is a nonconvex optimization problem, since the objective
is a convex function, and we wish to maximize it; we will explain below how we
can approximately solve it.


\subsection{Interpretation via a simple trading policy}\label{s-simple-policy}
While our formulation of the stat-arb optimization problem \eqref{e-stat-arb-prob}
makes sense on its own, we can further motivate it by looking at the profit 
obtained using a simple trading policy.
Suppose we hold quantity $q_t\in \reals$ of the portfolio, \ie, we hold
the portfolio $q_t s\in \reals^n$ (in shares).
We assume that $q_0=q_T=0$, \ie, we start and end with no holdings. In
period $t$ we buy $q_t-q_{t-1}$ and pay $p_t(q_t-q_{t-1})$. The total profit is then
\BEQ
\label{e-profit} \sum_{t=1}^{T-1} q_t (p_{t+1}-p_t). 
\EEQ 

We will relate this to our objective in \eqref{e-stat-arb-prob} above, 
with the simple trading policy
\BEQ
\label{e-simple-policy} q_t = \mu-p_t, \quad t=1, \ldots, T-1,
\EEQ 
which we refer to as a linear trading policy since the holdings are 
a linear function of the difference between the band midpoint and the current price.
This trading policy holds nothing when $p_t=\mu$, \ie, the price is 
in the middle of the band.
When the price is low,  $p_t = \mu-1$, we hold $q_t = +1$, and when it is
high, $p_t = \mu+1$ we hold $q_t = -1$. 

With the simple linear policy
\eqref{e-simple-policy} and the boundary conditions $q_0=q_T=0$, the 
profit \eqref{e-profit} is, after some algebra,
\[
\frac{1}{2}\sum_{t=2}^T (p_t-p_{t-1})^2 + \frac{
(p_1-\mu)^2 -(p_T-\mu)^2}{2}.
\]
The first term is one half our objective.  
The second term is between $-1/2$ and $1/2$, since $(p_T-\mu)^2$ and
$(p_1-\mu)^2$ are both between $0$ and $1$. Thus the profit is at least
\BEQ\label{e-lower-bound-profit}
\frac{1}{2}\left( \sum_{t=2}^T (p_t-p_{t-1})^2 - 1 \right).
\EEQ
This shows that our objective is the profit of the simple 
linear policy \eqref{e-simple-policy}, scaled by one-half, plus a 
constant.
In particular, if the objective of the problem \eqref{e-stat-arb-prob} exceeds one, 
the simple linear policy makes a profit.

\subsection{Solution method}
\paragraph{Convex-concave procedure.}
We solve the problem~\eqref{e-stat-arb-prob} approximately using sequential
convex programming, specifically the convex-concave procedure
\cite{shen2016disciplined, lipp2016variations}. Let $k$ denote the iteration, with $s^k$ the portfolio and $p^k_t$ the portfolio
price in the $k$th iteration. In each iteration of the convex-concave procedure,
we linearize the objective, replacing the quadratic function $f(p) =
\sum_{t=2}^T (p_{t} - p_{t-1})^2$ with the affine approximation
\[
\hat f(p; p^k) = f(p^k) + \nabla f(p^k)^T(p-p^k) =
\nabla f(p^k)^T p + c,
\]
where $c$ is a constant (\ie, does not depend on $p$). This linearization is a
lower bound on the true objective, \ie, we have $f(p) \geq \hat f(p; p^k)$ for
all $p$.  For completeness we note that
\[
(\nabla f(p))_t = 
\begin{cases}
   2(p_1-p_2) & t = 1\\
   2(2p_t - p_{t-1} - p_{t+1}) & t = 2,\ldots,T-1\\
   2(p_T-p_{T-1})& t = T.
\end{cases}
\]
We now solve the linearized problem 
\BEQ
\begin{array}{ll}
\mbox{maximize} & \hat f(p; p^k)\\
\mbox{subject to} & -1 \leq p_t - \mu \leq 1, \quad p_t=s^TP_t, \quad t = 1,
\ldots, T\\
&|s|^T \bar P \leq L, \quad \mu \geq 0,
\end{array}
\label{eq:vol_prob_scp}
\EEQ 
with variables $p_t$, $s$ and $\mu$.  This is a convex problem, in fact a
linear program (LP), and readily solved~\cite{BoV:04}. 
We take the solution of this problem as
the next iterate $p^{k+1}$, $s^{k+1}$, $\mu^{k+1}$.
This simple algorithm converges to a local solution of \eqref{e-stat-arb-prob},
typically in at most a few tens of iterations.

\paragraph{Cleanup phase.}
The leverage constraint $|s|^T \bar P \leq L$ encourages sparse solutions,
but in some cases the convex-concave procedure converges to a portfolio with 
a few small holdings.
To achieve even sparser portfolios, we can carry out a clean-up step once
the convex-concave procedure has converged.
We first determine the subset of assets for which $s_i$ is zero or small,
as measured by its relative weight in the portfolio, \ie, $i$ for which
\[
|s_i|\bar{P}_i \leq \eta |s|^T\bar{P},
\]
where $\eta$ is a small positive constant such as $0.05$.
We then solve the problem again, this time with the constraint
that all such $s_i$ are zero.
This takes just a few convex-concave iterations, and can be repeated,
which results in sparse portfolios in which every asset has weight
at least $\eta$.

\paragraph{Implementation.}
To make the optimization problem better conditioned, we scale the prices $P_t$
so that (after scaling) $\bar{P}=\ones$, the vector with all entries one.
Thus after scaling, the leverage
$|s|^T\bar{P}$ becomes the $\ell_1$ norm of $s$. We also scale the gradient (or
objective) to be on the order of magnitude one.
These scalings do not affect the solution, but make the method less vulnerable
to floating point rounding errors.

\paragraph{Initialization.}
The final portfolio found by the convex-concave procedure, plus the cleanup phase,
depends on the initial portfolio $s^1$.  It can and does converge to different 
final portfolios for different starting portfolios.
With a random initialization we can find multiple stat-arbs for the same
universe.  (We also get some duplicates when the method converges to the
same final portfolio from different initial portfolios.)
We have found that uniform initialization
of the entries of $s^1$ in the interval $[0,1]$ works well in practice.
Thus from one universe and data set, we can obtain multiple stat-arbs.

\section{Finding moving-band stat-arbs}
\label{s-movingband-statarbs}
\subsection{Moving-band stat-arbs}
In the fixed-band stat-arb problem~\eqref{e-stat-arb-prob} $\mu$ is constant, 
so the midpoint of the trading band does not vary with time.
In this section we describe a simple but powerful extension in which
the stat-arb band midpoint changes over time.
One simple (and traditional) choice is to define $\mu_t$ as the mean of 
the trailing prices $p_t$, for example the mean over the last $M$ periods,
\[
\mu_t = \frac{1}{M}\sum_{\tau=t-M+1}^t p_\tau.
\]
(This requires knowledge of the prices $P_0,P_{-1}, \ldots, P_{-M+1}$.)
In this formulation, $\mu_t$ is also a function of $s$, the portfolio,
but it is a known linear function of it.
Any other linear expression for the average recent price could be used, \eg,
exponentially weighted moving average (EWMA).
A moving-band stat-arb is a portfolio $s$ in which the price $p_t$ 
stays in a moving band with width two and midpoint $\mu_t$, and also has
high variance.

\paragraph{Trading policy.}
The simple linear trading policy~\eqref{e-simple-policy} can be modified in the
obvious way, as
\BEQ
\label{e-simple-policy-moving} 
q_t = \mu_t-p_t, \quad t=1, \ldots, T-1.
\EEQ 

\subsection{Finding moving-band stat-arbs}
We arrive at the optimization problem
\BEQ
\begin{array}{ll}
\mbox{maximize} & \sum_{t=2}^T (p_{t} - p_{t-1})^2\\
\mbox{subject to} & -1 \leq p_t - \mu_t \leq 1, \quad p_t=s^TP_t, \quad t = 1,
\ldots, T\\
& |s|^T \bar P \leq L, \quad \mu_t = (1/M)\sum_{\tau=t-M+1}^t p_\tau, \quad t = 1, \ldots, T,
\end{array}
\label{e-moving-band-stat-arb}
\EEQ 
with variables $s\in \reals^n$, $p_1, \ldots, p_T$, 
and $\mu_1, \ldots, \mu_T$. (The latter two sets of variables are simple
linear functions of $s$.)
In this problem we have an additional parameter $M$, the memory for 
the band midpoint.

Note that in the fixed-band stat-arb problem \eqref{e-stat-arb-prob},
$\mu$ is a scalar variable that we freely choose; in the moving-band stat-arb
problem \eqref{e-moving-band-stat-arb}, $\mu_t$ varies over time,
and is itself a function of $s$.
Despite this complication, the moving-band stat-arb problem 
\eqref{e-moving-band-stat-arb} can be (approximately) solved using
exactly the same convex-concave method as the fixed-band
stat-arb problem \eqref{e-stat-arb-prob}; the only difference is
in the convex constraints.



\section{Numerical experiments}\label{s-experiments}
We illustrate our method with an empirical study on historical asset prices. Everything needed to reproduce the results is
available online at
\begin{quote}
\centering
\url{https://github.com/cvxgrp/cvxstatarb}.
\end{quote}

\subsection{Experimental setup}
\paragraph{Data set.}
We use daily data of the CRSP US Stock Databases from the Wharton
Research Data Services (WRDS) portal~\cite{WRDS}. The data set consists of
adjusted prices of 15405 assets from January 4th, 2010, to 
December 30th, 2023, for a total of 3282 trading days.

\paragraph{Monthly search for stat-arbs.}
Starting December 23, 2011, and every 21 trading days thereafter, until July
6th, 2022, we use the convex-concave method with 10 different random 
initial portfolios.
From these 10 stat-arbs we add the unique ones, defined
by the set of assets in the stat-arb, to our current set of stat-arbs. 
All together we solve fixed-band and moving-band problems 1270 times.

\paragraph{Parameters.}
We set $L=\$50$
for the fixed-band stat-arb and $L=\$100$ for the moving-band stat-arb.
For the moving-band stat-arb, we take $M=21$ days, \ie, we use the trailing 
month average price as the midpoint,
a value commonly used for Bollinger
bands~\cite{bollinger1992using, bollinger2002bollinger}.
We note that our results are not very sensitive to the choices of $L$ and $M$, and 
that choosing a
larger $L$ for the moving-band than for the fixed-band stat-arb is reasonable
since we expect the price of a portfolio to vary less around its short-term
moving midpoint than around a fixed midpoint.

\paragraph{Trading policy.}
We use the simple linear trading policies defined in~\eqref{e-simple-policy} and 
\eqref{e-simple-policy-moving} for the fixed-band and moving-band stat-arbs,
respectively. 
%
We use a simple time-based exit condition, where we trade a stat-arb for
$T^{\max}$ trading days, and then
exit the position uniformly (linearly) over the next
$T^{\text{exit}}$ trading days.  This means we take 
\[
q_t = (1-\alpha_t)(\mu-p_t), \quad 
t=T^{\max},\ldots,T^{\max}+T^{\text{exit}}-1,
\]
where $\alpha_t = (t+1-T^{\max})/T^{\text{exit}}$.
We use parameter values
$T^{\max}=63$ for the fixed-band stat-arb and $T^{\max}=125$ for the moving-band stat-arb,
and $T^\text{exit}=21$ for both.
We also exit a stat-arb if the value of the stat-arb plus a cash account drops below
a given level, as described below.

\subsection{Simulation and metrics}
We simulate and evaluate a stat-arb as follows. 

\paragraph{Cash account.}
Each stat-arb is initialized with a cash account value
\[
C_0 = \nu |s|^T P_0,
\]
where $\nu$ is a positive parameter and $P_0$ is the price vector the day before we
start trading the stat-arb. 
We take $\nu=0.5$ in our experiments. The cash account is updated as
\[
C_{t+1} = C_t - (q_{t+1}-q_t)p_{t+1} - \phi_t, \quad t=1,\ldots, T,
\]
where $\phi_t$ is the transaction and holding
cost, consisting of the trading cost at time $t+1$ and the holding cost
over period $t$, described below. 
The cash account plus the long position is meant to be the collateral for 
the short positions.

\paragraph{Portfolio net asset value.}
The net asset value (NAV) of the portfolio, including the cash
account, at time $t$ is then 
\[
V_{t} = C_t + q_t p_t.
\]
(Note that $V_{0} = C_0$.) The profit at time $t$ is
\[
V_t - V_{t-1} = q_tp_t +C_t - q_{t-1}p_{t-1} - C_{t-1} = q_t(p_t-p_{t-1}) - \phi_t,
\]
which agrees with the profit formula~\eqref{e-profit}, after accounting for
transaction and holding costs.

\paragraph{NAV based termination.}
If the NAV goes below 50\% of $C_0$, we liquidate the stat-arb portfolio. 
We do this for two reasons. First, we do not want the portfolio to have negative
value, \ie, to go bust. 
Second, this constraint ensures
that the short positions are always fully collateralized by the long positions
plus the cash account, with a margin of at least half of our initial investment.
All of our metrics include early termination stat-arbs.

\paragraph{Trading and shorting costs.} Our numerical experiments take into account 
transaction costs, \ie, we buy assets at the ask price, 
which is the (midpoint) price
plus one-half the bid-ask spread, and we sell assets at the bid price,
which is the price minus one-half the bid-ask spread.
Note that while we do not take into account
transaction cost in our simple trading policy, we do in simulation and 
accounting.
We use 0.5\% as a proxy for
the annual shorting cost of assets, which is well above what is typically
observed in practice for liquid assets~\cite{d2002market, geczy2002stocks,
kim2023shorting, drechsler2014shorting}. (We also note that the method is robust against shorting costs; the results are consistent even when shorting costs exceed 10\% per annum.)

\paragraph{Metrics.}
The profit of a stat-arb is 
\[ 
\sum_{t=1}^{T}{(V_t - V_{t-1})},
\]
where $T=T^{\max} + T^{\text{exit}}$ is the number of trading days in the
evaluation period. The return at time $t$ is 
\[
r_t = \frac{V_t - V_{t-1}}{V_{t-1}}, \quad t = 1, \ldots, T.
\]
We report several
standard metrics based on the daily returns $r_t$. The average return is 
\[
\overline r = \frac{1}{T} {\sum_{t=1}^{T}{r_t}},
\]
which we multiply by $250$ to annualize.
The risk (return volatility) is 
\[
\left(\frac{1}{T} \sum_{t=1}^{T}\left(r_t-\overline r\right)^2\right)^{1/2},
\]
which we multiply by $\sqrt{250}$ to annualize.
The annualized Sharpe ratio is the ratio of the annualized average return to the
annualized risk. Finally, the maximum drawdown is
\[
\max_{1 \leq t_1 < t_2\leq T} \left( \frac{V_{t_1}}{V_{t_2}} - 1 \right),
\]
the maximum drop in value form a previous high.

\subsection{Results for fixed-band stat-arbs}\label{s-results}
\begin{figure}
\centering
\includegraphics[width=0.6\textwidth]{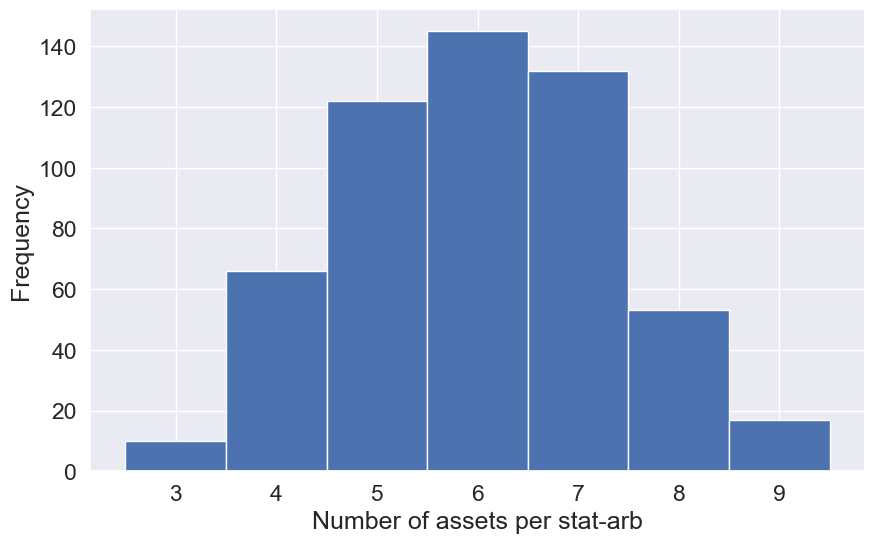}
\caption{Distribution of the number of assets per fixed-band stat-arb.}
\label{f-n-assets-constant}
\end{figure}

\begin{figure}
\centering
\includegraphics[width=0.6\textwidth]{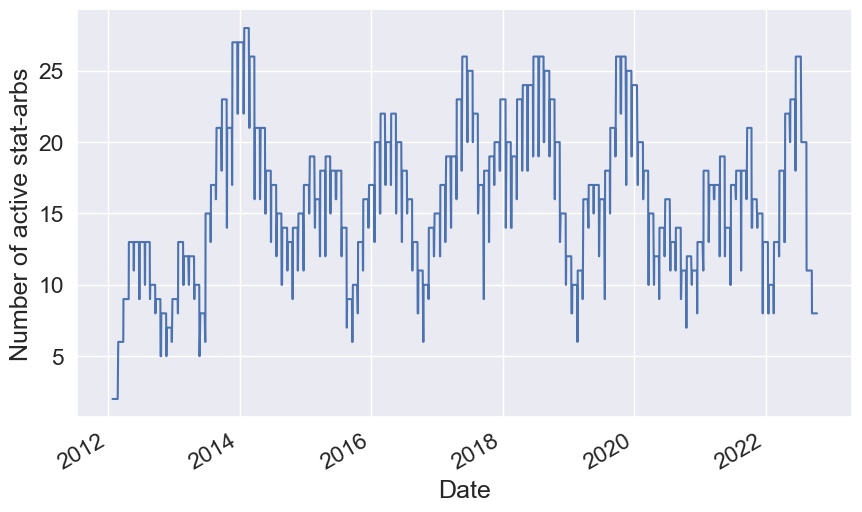}
\caption{Number of active fixed-band stat-arbs over time.}
\label{f-n-active-constant}
\end{figure}

\paragraph{Stat-arb statistics.} 
After solving the fixed-band stat-arb problem 1270 times, we found
545 unique stat-arbs.  These stat-arbs contained between $3$ and $9$ assets,
with a median value of $6$, 
as shown in figure~\ref{f-n-assets-constant}. 
Over time the number of active stat-arbs ranges up to 28, with a median 
value of 17, as shown in figure~\ref{f-n-active-constant}.
Out of the 545 stat-arbs, 26 (around 5\%) were terminated
before the end of the evaluation period, due to the NAV falling below 
50\% of the initial investment.

\paragraph{Metrics.}
Table~\ref{t-profitability-constant} summarizes metrics related to the profitability of the
fixed-band stat-arbs. 
Of the 545 stat-arbs, 63\% were profitable.  The average annualized return is
10\%, with an average annualized risk of 32\%, and an average annualized Sharpe
ratio of 0.81. The maximum drawdown was on average 15\% over the four-month
trading period for each stat-arb. 

\begin{table}
\centering
\begin{tabular}{lr}
\midrule
\multicolumn{2}{l}{\textbf{Profitability}} \\
Fraction of profitable stat-arbs & 70\% \\
\midrule
\multicolumn{2}{l}{\textbf{Annualized return}} \\
Average & 10\% \\
Median & 18\% \\
75th percentile & 33\% \\
25th percentile & -3\% \\
\midrule
\multicolumn{2}{l}{\textbf{Annualized risk}} \\
Average  & 32\% \\
Median  & 21\% \\
75th percentile  & 36\% \\
25th percentile   & 12\% \\
\midrule
\multicolumn{2}{l}{\textbf{Annualized Sharpe ratio}} \\
Average  & 0.79 \\
Median  & 1.04 \\
75th percentile  & 1.78 \\
25th percentile   & -0.08 \\
\midrule
\multicolumn{2}{l}{\textbf{Maximum drawdown}} \\
Average  & 15\% \\
Median  & 10\% \\
75th percentile  & 19\% \\
25th percentile   & 5\% \\
\midrule
\end{tabular}
\caption{Metric summary for 545 fixed-band stat-arbs.}
\label{t-profitability-constant}
\end{table}

\begin{figure}
\centering
\includegraphics[width=0.6\textwidth]{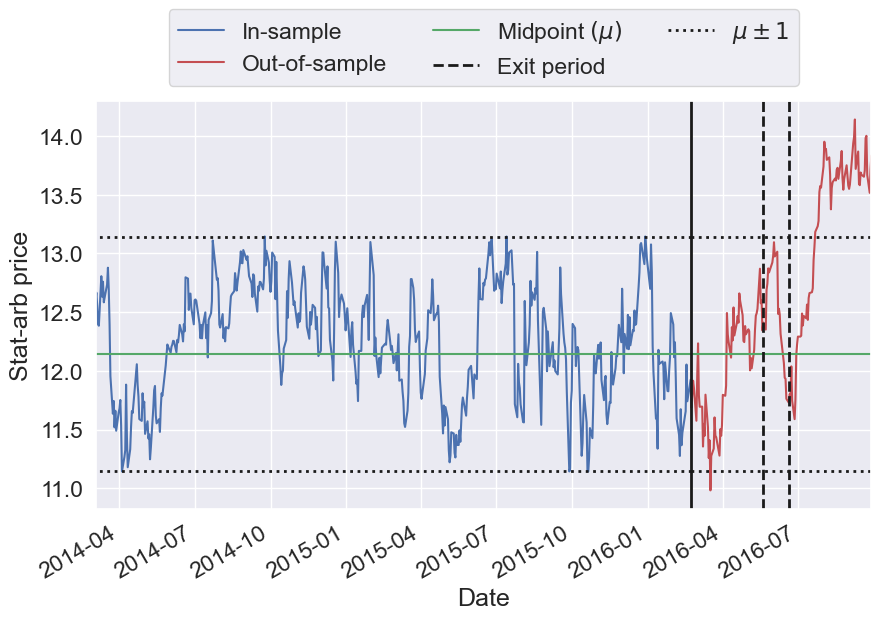}
\includegraphics[width=0.6\textwidth]{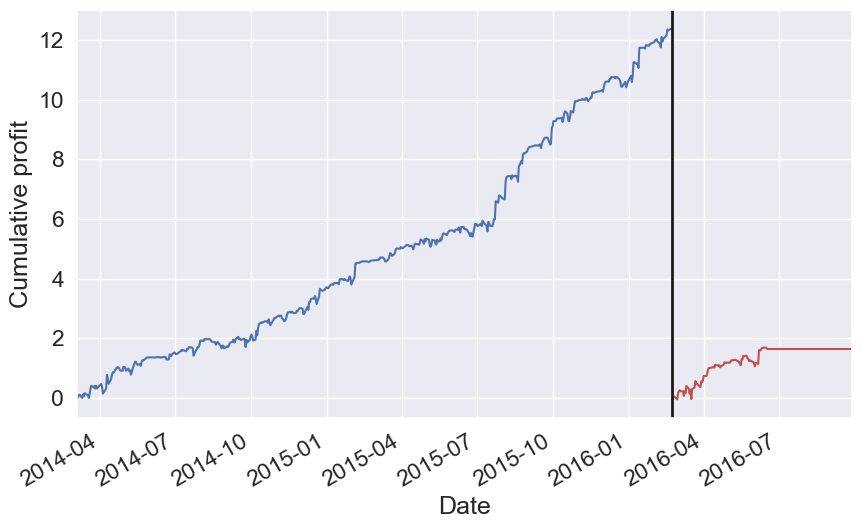}
\caption{A fixed-band stat-arb strategy that made money.  \emph{Top.} Price.
\emph{Bottom.} Cumulative profit.}
\label{f-SA-const-win}
\end{figure}

\begin{figure}
\centering
\includegraphics[width=0.6\textwidth]{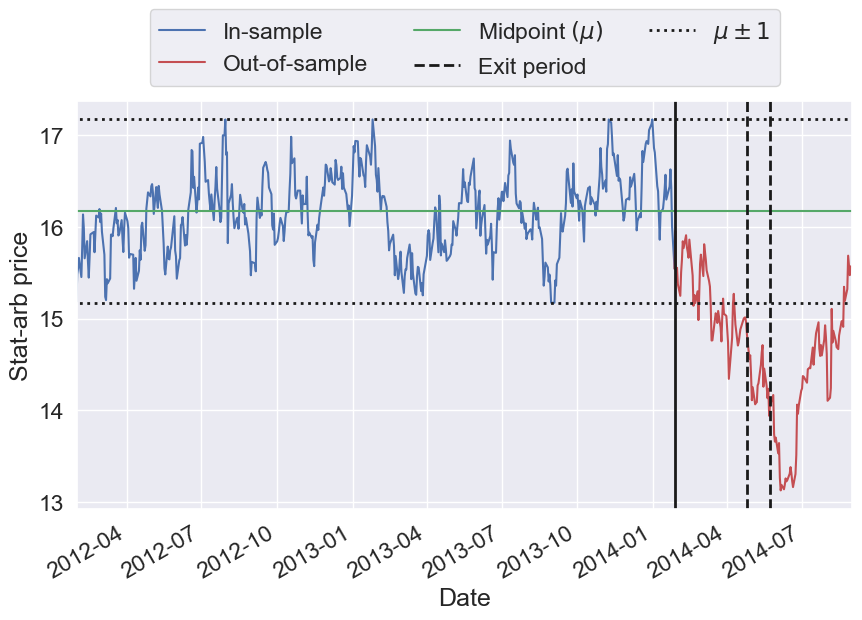}
\includegraphics[width=0.6\textwidth]{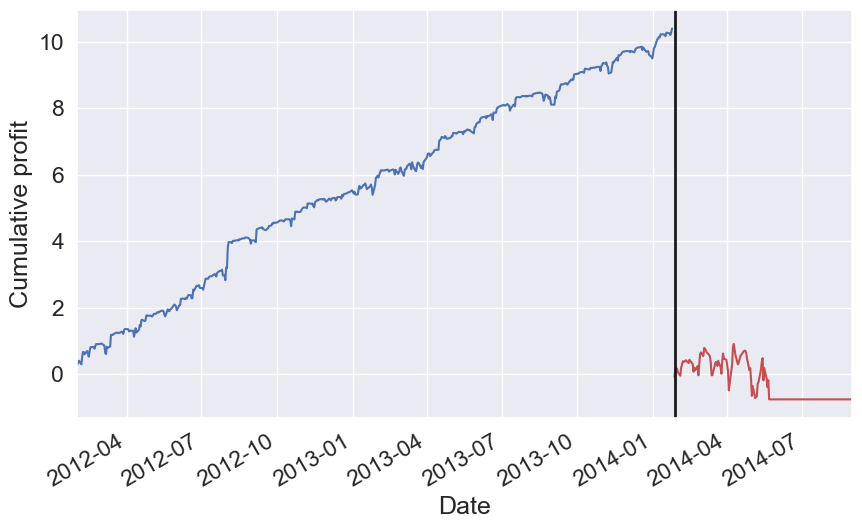}
\caption{
A fixed-band stat-arb strategy that lost money.
\emph{Top.} Price.  \emph{Bottom.} Cumulative profit.}
\label{f-SA-const-loss}
\end{figure}

\clearpage
\paragraph{Example stat-arbs.}
We show the detailed evolution of two stat-arbs,
one that made money and one that lost money, in figures
\ref{f-SA-const-win} and \ref{f-SA-const-loss}, respectively. These were chosen
to have average returns around the 75th and 25th percentiles of the return
distribution across our 545 stat-arbs.
As expected both stat-arbs are very profitable in-sample. 
The first one continues to be profitable out-of-sample.

The first one, which made money, contained the assets
\begin{center}
\begin{tabular}{l}
Facebook\\
The Walt Disney Company\\
Eli Lilly and Company\\
Biogen\\
Occidental Petroleum \\
Alexion Pharmaceuticals.
\end{tabular}
\end{center}
The second one, which lost money, contained the assets
\begin{center}
\begin{tabular}{l}
Coca Cola \\
Bristol-Myers Squibb \\
eBay\\
Walgreens \\
ArcelorMittal \\
Valero Energy
\end{tabular}
\end{center}

\clearpage
\subsection{Results for moving-band stat-arbs}\label{s-results-moving}
\begin{figure}
\centering
\includegraphics[width=0.6\textwidth]{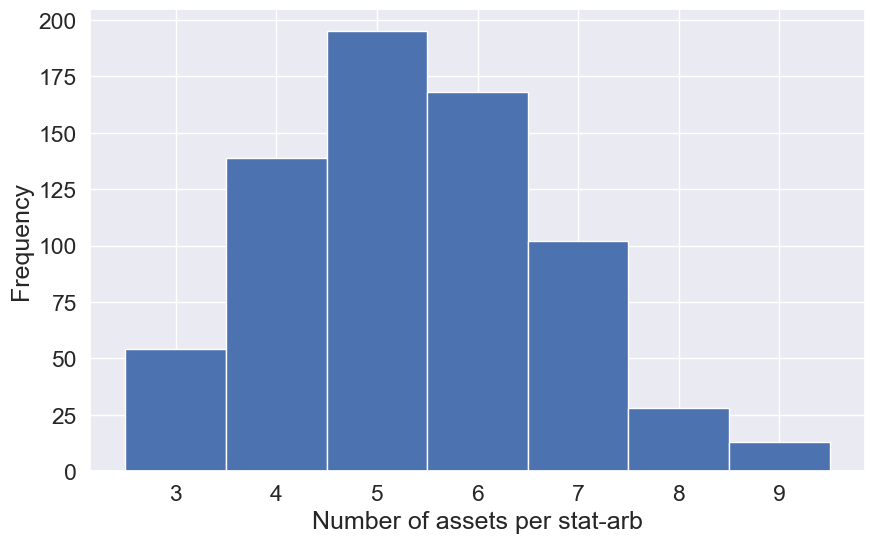}
\caption{Distribution of the number of assets per moving-band stat-arb.}
\label{f-n-assets-moving}
\end{figure}

\begin{figure}
\centering
\includegraphics[width=0.6\textwidth]{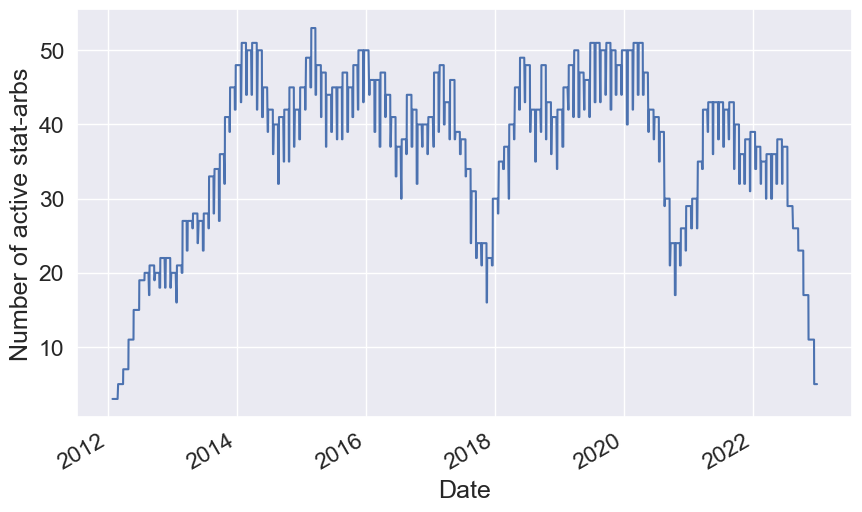}
\caption{Number of active moving-band stat-arbs over time.}
\label{f-n-active-moving}
\end{figure}

\paragraph{Stat-arb statistics.}
We found 712 unique moving-band stat-arbs (compared with 545 fixed-band stat-arbs).
These stat-arbs contained between 1 and 10 assets, with a median value of 5.
The full distribution is shown in figure~\ref{f-n-assets-moving}.
The number of active moving-band stat-arbs over time is shown in
figure~\ref{f-n-active-moving}.
The median number of active stat-arbs is 40. 
This is considerably larger than the number of active fixed-band stat-arbs since
we find more of them, and they are active (by our choice) almost twice as long.
Only three (around 0.4\%) out of the 712 moving-band stat-arbs were 
terminated before the end of the evaluation period, due to the NAV falling below 
50\% of the initial investment. 


\paragraph{Metrics.} 
Table~\ref{t-profitability-moving} summarizes the profitability of the
moving-band stat-arbs.
A large majority (70\%) of the stat-arbs are profitable. The average annualized
return was 15\%, with an average annualized risk of 20\%, and an 
average annualized Sharpe ratio of 0.84. The maximum drawdown was on average
12\% over the seven-month trading period for each stat-arb. 

\begin{table}
\centering
\begin{tabular}{lr}
\midrule
\multicolumn{2}{l}{\textbf{Profitability}} \\
Fraction of profitable stat-arbs & 70\% \\
\midrule
\multicolumn{2}{l}{\textbf{Annualized return}} \\
Average & 15\% \\
Median & 12\% \\
75th percentile & 24\% \\
25th percentile & 3\% \\
\midrule
\multicolumn{2}{l}{\textbf{Annualized risk}} \\
Average  & 20\% \\
Median  & 15\% \\
75th percentile  & 25\% \\
25th percentile   & 9\% \\
\midrule
\multicolumn{2}{l}{\textbf{Annualized Sharpe}} \\
Average  & 0.84 \\
Median  & 0.88 \\
75th percentile  & 1.52 \\
25th percentile   & 0.21 \\
\midrule
\multicolumn{2}{l}{\textbf{Maximum drawdown}} \\
Average  & 12\% \\
Median  & 9\% \\
75th percentile  & 15\% \\
25th percentile   & 5\% \\
\midrule
\end{tabular}
\caption{Metric summary for 712 moving-band stat-arbs.}
\label{t-profitability-moving}
\end{table}

\paragraph{Comparison with fixed-band stat-arbs.}
Our first observation is that far fewer of the moving-band stat-arbs were
terminated early due to low NAV than the fixed-band stat-arbs, despite their
running for a period almost twice as long.
Comparing tables~\ref{t-profitability-constant} and \ref{t-profitability-moving} 
we see that the metrics for fixed-band
stat-arbs are more variable, with a larger range in each of the metrics. 
The moving-band stat-arbs are more profitable than the fixed-band stat-arbs, 
but the difference is not large.

\begin{figure}
   \centering
   \includegraphics[width=0.6\textwidth]{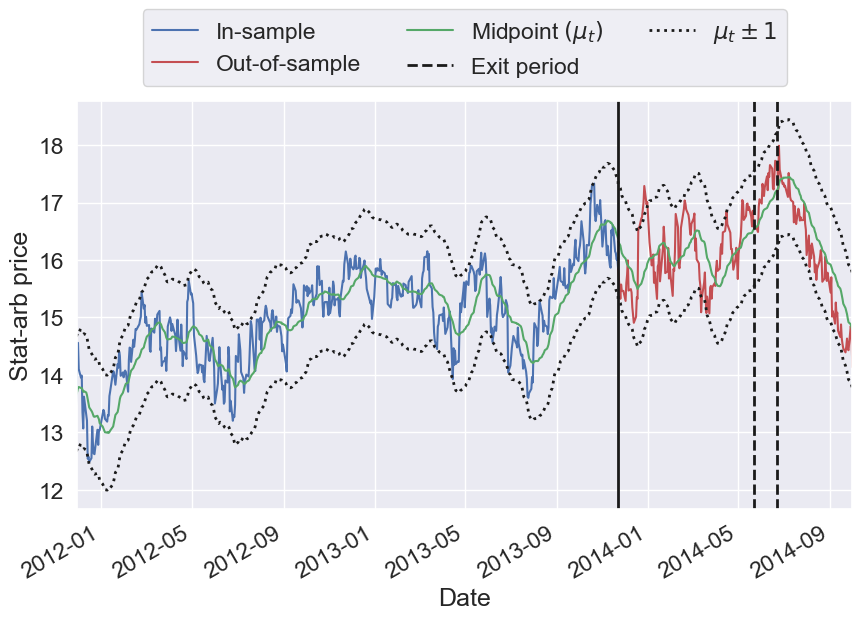}
   \includegraphics[width=0.6\textwidth]{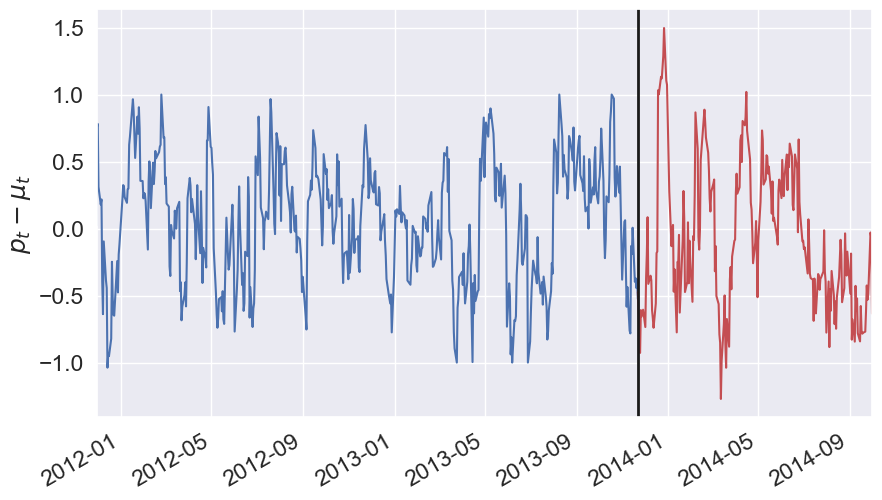}
   \includegraphics[width=0.6\textwidth]{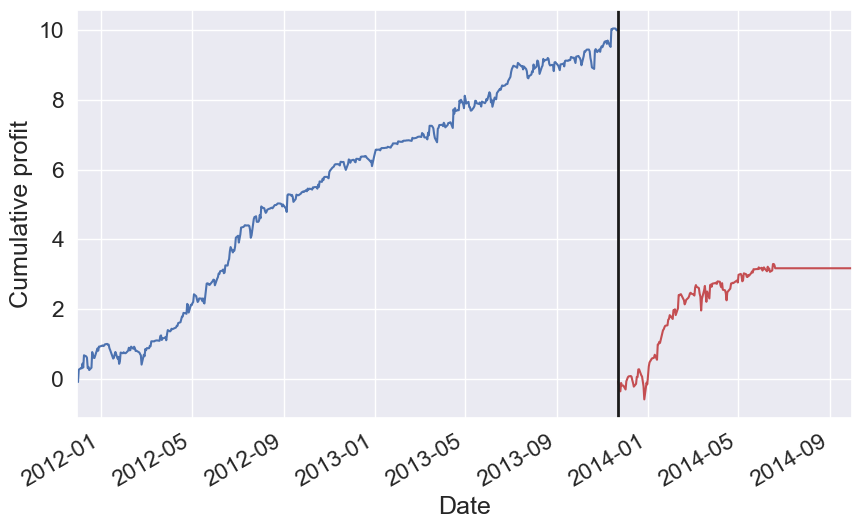}
   \caption{A moving-band stat-arb strategy that made money.
   \emph{Top.} Price. \emph{Middle.} Price
   relative to the trailing mean.  \emph{Bottom.} Cumulative profit.}
   \label{f-SA-moving-win}
   \end{figure}

\begin{figure}
\centering
\includegraphics[width=0.6\textwidth]{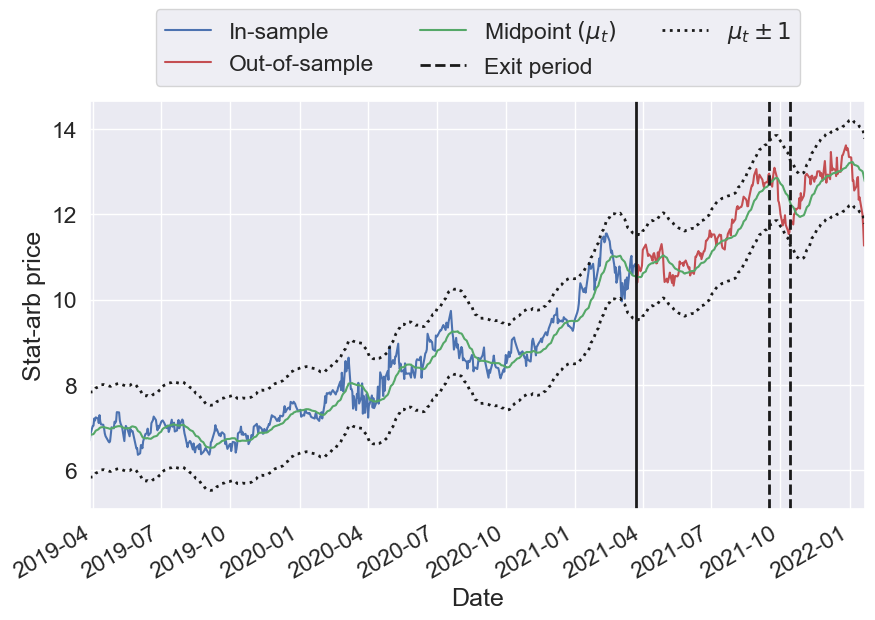}
\includegraphics[width=0.6\textwidth]{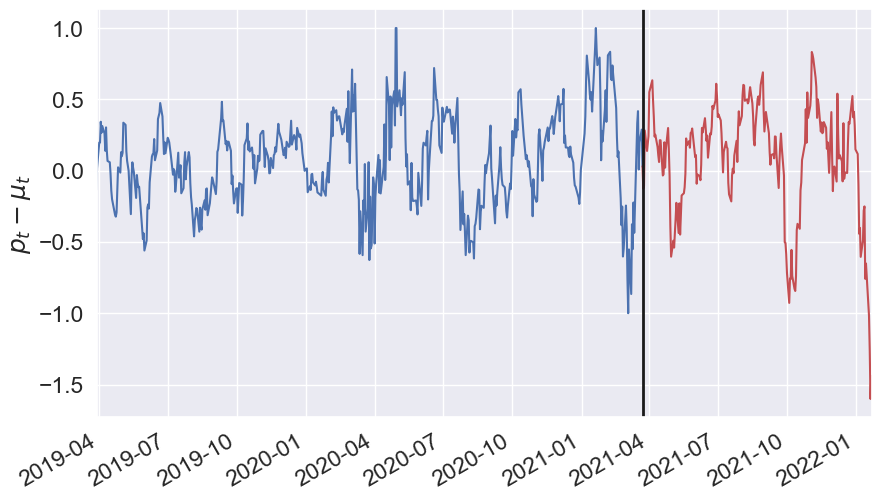}
\includegraphics[width=0.6\textwidth]{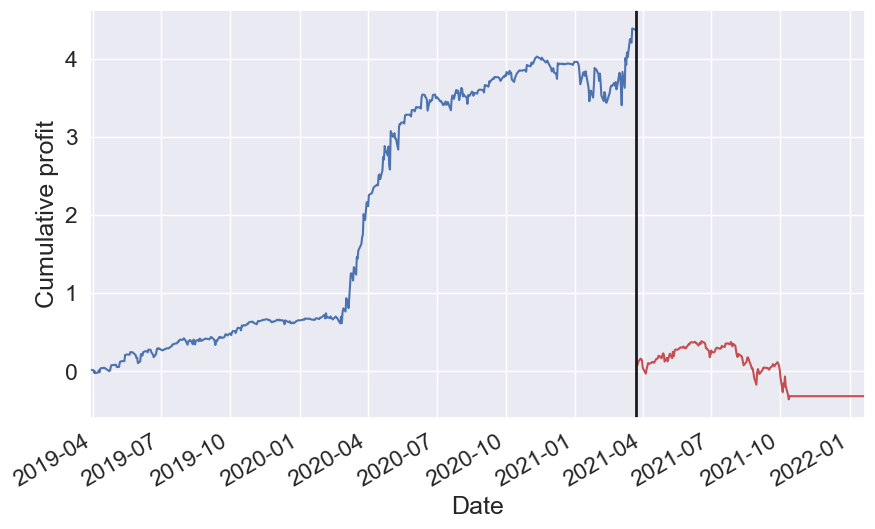}
\caption{A moving-band stat-arb strategy that lost money. 
\emph{Top.} Price. \emph{Middle.} Price
relative to the trailing mean. \emph{Bottom.} Cumulative profit.}
\label{f-SA-moving-loss}
\end{figure}

\paragraph{Example stat-arbs.}
Two stat-arbs, picked to represent roughly the 70th and 15th percentiles of the return
distribution across the 712 stat-arbs, are illustrated in figures~\ref{f-SA-moving-win} and
\ref{f-SA-moving-loss}, respectively.
Again, both stat-arbs are profitable in-sample, and the first one continues to
be profitable out-of-sample.

The first one, which made money, contained the assets
\begin{center}
\begin{tabular}{l}
Morgan Stanley \\
Monsanto\\
Walgreens \\
Accenture Plc \\
ArcelorMittal \\
Pioneer Natural Resources
\end{tabular}
\end{center}
The second one, which lost money, contained the assets
\begin{center}
\begin{tabular}{l}
Lockheed Martin \\
ServiceNow \\
Gilead Sciences \\
NXP Semiconductors NV
\end{tabular}
\end{center}

\clearpage

\section{Conclusions and comments}\label{s-conclusion}

We have formulated the problem of finding stat-arbs
as a nonconvex optimization problem which can be approximately solved
using the convex-concave procedure.
We have introduced moving-band stat-arbs, which combine ideas from 
statistical arbitrage and price band trading.

Our empirical study on historical data shows that moving-band 
stat-arbs perform better than fixed-band stat-arbs, and remain 
profitable for longer out-of-sample periods.  
Our empirical study uses very simple trading and exit policies;
we imagine that with more sophisticated ones such as those cited above,
the results would be even better.
Our focus in this paper is on finding stat-arbs, and not on trading them.

\paragraph{Variations and extensions.}
We mention here several ideas that we tried out, but were surprised to find 
did not improve the empirical results.
\BIT
\item \emph{Asset screening.} We construct stat-arbs using assets only 
within an industry or sector.
\item \emph{Validation.} We split past asset prices into a training and a 
test set.  We find candidate stat-arbs using the training data and then test them
on the test data. We then only trade those with good test performance.
\item \emph{Incorporating transaction costs in the trading policy.}
We modify the linear policy to take into account transaction costs.
(Our simulations take trading cost into account, but our simple
linear trading policy does not.)
\item \emph{Hysteresis-based trading.}  We use a hysteresis-based trading policy,
which can help reduce transaction costs compared to the linear policy.
\EIT

\paragraph{Trading a portfolio of stat-arbs.}
We have focussed on finding individual stat-arbs.  A next obvious topic is
how to trade a portfolio of stat-arbs. This will be addressed in an upcoming
paper by the authors.  We simply note here that the results presented in this
paper are all for single stat-arbs, and so fall somewhere in between
individual assets and a full portfolio.

\subsection*{Acknowledgments}
The idea of using the convex-concave procedure to find fixed-band stat-arbs is
from initial unpublished work by Stephen Boyd and Jonathan Tuck.
We thank Ron Kahn and Mark Mueller for detailed reviews of the paper and many
helpful comments and suggestions. We gratefully acknowledge support from the Office
of Naval Research. This work was partially supported by ACCESS -- AI Chip Center for Emerging Smart Systems.
Kasper Johansson was partially funded by the Sweden America Foundation.

\clearpage
\bibliography{refs}

\end{document}